\newcommand{\resection}[1]{\setcounter{equation}{0}\section{#1}}

\documentclass[a4paper,11pt]{article}
\usepackage{amsfonts}

\usepackage{amsmath}

\newcommand{\EQ}{\begin{equation}}
\newcommand{\EN}{\end{equation}}
\newcommand{\bea}{\begin{eqnarray}}
\newcommand{\eea}{\end{eqnarray}}
\newcommand{\hs}{\hspace{0.1cm}}

\begin{document}

\setcounter{page}{0} \topmargin0pt \oddsidemargin5mm \renewcommand{%
\thefootnote}{\arabic{footnote}}\newpage \setcounter{page}{0} 
\begin{titlepage}
\begin{flushright}
SISSA 36/2008/EP
\end{flushright}
\vspace{0.5cm}
\begin{center}
{\large{\bf On the space of quantum fields in massive two-dimensional theories
}
}\\

\vspace{1.8cm}
{\large Gesualdo Delfino
} 
\\
\vspace{0.5cm}
{\em International School for Advanced Studies (SISSA)}\\
{\em via Beirut 2-4, 34014 Trieste, Italy}\\
{\em INFN sezione di Trieste}\\
\end{center}
\vspace{1.2cm}

\renewcommand{\thefootnote}{\arabic{footnote}}
\setcounter{footnote}{0}

\begin{abstract}
\noindent
For a large class of integrable quantum field theories we show that the 
$S$-matrix determines a space of fields which decomposes into subspaces
labeled, besides the charge and spin indices, by an integer $k$. For scalar
fields $k$ is non-negative and is naturally identified as an off-critical
extension of the conformal level. To each particle we associate an operator
acting in the space of fields whose eigenvectors are primary ($k=0$) fields
of the massive theory. We discuss how the existing results for models as 
different as $Z_n$, sine-Gordon or Ising with magnetic field fit into this 
classification.
\end{abstract}

\end{titlepage}

\newpage

\resection{Introduction}
The classification of quantum fields according to the scaling dimension 
determined by the short distance behavior of correlation functions is at 
the basis of the renormalization group idea and its many applications to 
physics. The field theoretical description of critical phenomena assumes
the existence of a spectrum of scaling dimensions bounded from below,
with the lowest dimension (primary) fields determining the critical 
exponents and the others responsible for corrections to scaling. Such a
scheme is expected to hold in the generic case of non-trivial fixed points 
and also in absence of reflection positivity, when the scaling dimensions 
are not necessarily positive. 

An exact confirmation exists for the fixed point (conformal) field theories 
in two dimensions \cite{BPZ}. There, remarkably, the semi-infinite spectrum 
of scaling dimensions emerges in connection with the lowest weight 
representations of the  infinite-dimensional conformal group. It is found
that each representation corresponds to a family of fields whose scaling
dimensions differ by integers. For scalar fields this gradation according 
to the dimension can be made in terms of a single non-negative integer
called level.

Two-dimensional quantum field theory allows also for $massive$ exactly
solvable (integrable) theories \cite{ZZ}. Here, however, the solution comes 
in the form of an exact $S$-matrix, and the characterization of the space of 
fields starting only from particle dynamics, important in principle as 
well as for applications to off-critical systems, is a 
non-trivial task. A set of functional equations \cite{KW,Smirnovbook} is 
known for the matrix elements of fields on particle states (form factors). 
The space of fields of the massive theory corresponds to the space of
solutions of the form factor equations. These, however, contain only the
symmetry data (charges and spin) of the field, and nothing about its
scaling dimension. With time, evidence of a relation between the high
energy asymptotics of form factors and the conformal levels emerged.
The first result in this sense was obtained in \cite{CM}, but only recently
the isomorphism between critical and off-critical field spaces has been
shown for the simplest massive model originating from a non-trivial
fixed point of the renormalization group \cite{isom}. 

The emergence of a sufficiently general pattern, however, had been so far
prevented by the difficulty of controlling the role of model dependent 
features such as the particle spectrum, the symmetries and the properties 
of mutual locality between particles and fields.
In this paper we show that the space of fields admits an unifying
characterization within a large class of integrable theories,
those with reflectionless scattering at high energies and additive charges. 
Starting from the $S$-matrix, the space of fields can be 
decomposed into subspaces labeled by four indices. Three of them encode the 
internal and Lorentz symmetry properties of the field, while the fourth is 
an integer, let us call it $k$, 
related to the asymptotic behavior of form factors. We obtain a lower 
bound for the asymptotic behavior which amounts to a lower bound on
$k$. For scalar fields $k$ is non-negative and is naturally identified as
playing the same role as the level in the conformal classification. 
We also associate to each particle species 
$a$ an operator $\Lambda_a$ mapping fields into fields
and whose "eigenvectors" are primary ($k=0$) fields. We show how 
this scheme applies to models (${\bf Z}_N$, sine-Gordon, the theories 
without internal symmetries) which are usually seen as sharing 
little else than integrability. 

The paper is organized as follows. In the next section we recall the
features of factorized scattering theories and, after restricting to the
subclass of interest in this paper, we give a characterization of the 
associated space of fields in terms of spin and charges. In section~3
we introduce the operators $\Lambda_a$, further classify the fields 
according to the asymptotic behavior of form factors  and discuss the 
analogies with the conformal classification. In section~4 we specialize the 
discussion to several models before making few concluding remarks
in section~5.

\resection{Integrable quantum field theories}
Integrable quantum field theories are characterized by a completely elastic
and factorized $S$-matrix \cite{ZZ}. If we denote by $A_a(\theta)$ a particle
of species $a$, with mass $m_a$ and energy-momentum $(p^0,p^1)=(m_a\cosh\theta,
m_a\sinh\theta)$, the $S$-matrix is completely determined by the two-particle
scattering amplitudes associated to the processes
\EQ
|A_a(\theta_1)A_b(\theta_2)\rangle_{in}=S_{ab}^{cd}(\theta_1-\theta_2)
|A_c(\theta_1)A_d(\theta_2)\rangle_{out}\,.
\EN
The theories we consider are invariant under charge conjugation, space and
time reversal, so that the amplitudes satisfy
\bea
&& S_{ab}^{cd}(\theta)=S_{\bar{a}\bar{b}}^{\bar{c}\bar{d}}(\theta)\,,\\
&& S_{ab}^{cd}(\theta)=S_{ba}^{dc}(\theta)\,,\\
&& S_{ab}^{cd}(\theta)=S_{cd}^{ab}(\theta)\,,
\eea
where indices with a bar denote anti-particles.

The scattering amplitudes satisfy the unitarity, crossing, bootstrap and 
factorization equations
\EQ
S_{ab}^{ef}(\theta)S_{ef}^{cd}(-\theta)=\delta_a^c\delta_b^d\,,
\label{unitarity}
\EN
\EQ
S_{ab}^{cd}(i\pi-\theta)=S_{a\bar{d}}^{c\bar{b}}(\theta)\,,
\label{crossing}
\EN
\EQ
\Gamma_{ab}^cS_{dc}^{ef}(\theta)=\Gamma_{hj}^fS_{da}^{ih}(\theta-i\bar{u}_{a
\bar{c}}^{\bar{b}})S_{ib}^{ej}(\theta+i\bar{u}_{b\bar{c}}^{\bar{a}})\,,
\label{bootstrap}
\EN
\EQ
S_{ab}^{de}(\theta_1-\theta_2)S_{dc}^{fg}(\theta_1-\theta_3)
S_{eg}^{hi}(\theta_2-\theta_3)=
S_{bc}^{de}(\theta_2-\theta_3)S_{ae}^{gi}(\theta_1-\theta_3)
S_{gd}^{fh}(\theta_1-\theta_2)\,,
\label{yb}
\EN
where repeated indices are summed over, $iu_{ab}^c$ is the resonant rapidity 
difference associated to the bound state formation $A_aA_b\to A_c$, 
$\bar{u}_{ab}^c\equiv\pi-u_{ab}^c$, and the three particle couplings 
$\Gamma_{ab}^c$ are determined by
\EQ
\mbox{Res}_{\theta=iu_{ab}^c}S_{ab}^{de}(\theta)=i\Gamma_{ab}^c\Gamma_{\bar{d}
\bar{e}}^{\bar{c}}\,.
\EN
The above equations are normally sufficient for the exact determination of the
scattering amplitudes.

In this paper we consider integrable theories invariant under the action of an 
abelian group of transformations $G$,  whose $S$-matrix
becomes diagonal at high energies, i.e. satisfies
\EQ
\lim_{\theta\to\pm\infty}S_{ab}^{cd}(\theta)=e^{\pm 2i\pi\alpha_{ab}}\delta_a^c
\delta_b^d\,;
\label{diagonal}
\EN
to be definite, in the following we take $\alpha_{ab}\in[0,1)$.

In the integrable quantum field theories the fields $\Phi(x)$ are constructed 
from the knowledge of the $S$-matrix through the determination of the form 
factors 
\begin{equation}
F_{a_1\ldots a_n}^{\Phi}(\theta _{1},\ldots,\theta_{n})=\langle 0|\Phi (0)|
A_{a_1}(\theta_{1})\ldots A_{a_n}(\theta_{n})\rangle\,,  
\label{ff}
\end{equation}
where $|0\rangle$ denotes the vacuum (i.e. zero-particle) state.  If $C_a$ and 
$C_\Phi$ denote the charges of $A_a$ and $\Phi$ with respect to the group $G$, 
the form factors (\ref{ff}) vanish by symmetry unless 
\EQ
\sum_{j=1}^nC_{a_j}=-C_\Phi\,.
\label{neutrality}
\EN
The form factors satisfy the equations \cite{KW,Smirnovbook}
\begin{eqnarray}
&&F_{a_1\ldots a_n}^{\Phi}(\theta_{1}+\Lambda,\ldots,\theta _{n}+\Lambda)=
e^{s_{\Phi}\Lambda }F_{a_1\ldots a_n}^{\Phi}(\theta_{1},\ldots,\theta_{n})\,,
\label{fn0} \\
&&F_{\ldots a_ia_{i+1}\ldots}^{\Phi}(\ldots,\theta _{i},\theta _{i+1},\ldots)=
S_{a_ia_{i+1}}^{b_ib_{i+1}}(\theta_{i}-\theta_{i+1})\,
F_{\ldots b_{i+1}b_i\ldots}^{\Phi}(\ldots,\theta_{i+1},\theta_{i},\ldots)\,,
\label{fn1} \\
&&F_{a_1\ldots a_n}^{\Phi }(\theta _{1}+2i\pi,\theta_{2},\ldots,\theta_{n})=
e^{-2i\pi\gamma_{\Phi,a_1}}F_{a_2\ldots a_na_1}^{\Phi}(\theta _{2},\ldots,
\theta_{n},\theta_{1})\,,
\label{fn2} \\
&&\mbox{Res}_{\theta_a-\theta_b=iu_{ab}^c}\,F_{aba_1\ldots a_n}^{\Phi}
(\theta_a,\theta_b,\theta_{1},\ldots,\theta_{n})=i\Gamma_{ab}^c\,
F_{ca_1\ldots a_n}^{\Phi}(\theta_c,\theta_{1},\ldots,\theta_{n})\,,
\label{fn3}\\
&&\mbox{Res}_{\theta^{\prime}=\theta+i\pi}\,F_{\bar{a}aa_1\ldots a_n}^{\Phi}
(\theta^{\prime},\theta,\theta_{1},\ldots,\theta _{n})=\nonumber\\
&&\hspace{1.5cm}i\left[
\delta_{a_1}^{b_1}\ldots\delta_{a_n}^{b_n}-e^{2i\pi\gamma_{\Phi,a}}
S_{a_1\ldots a_n}^{b_1\ldots b_n}(\theta|\theta_1,\ldots,\theta_n)\right] 
F_{b_1\ldots b_n}^{\Phi}(\theta_{1},\ldots ,\theta _{n})\,,  
\label{fn4}
\end{eqnarray}
where $s_\Phi$ is the euclidean spin of the field $\Phi(x)$,
\EQ
S_{a_1\ldots a_n}^{b_1\ldots b_n}(\theta|\theta_1,\ldots,\theta_n)\equiv
S_{a_1\alpha_n}^{b_1\alpha_1}(\theta-\theta_1)
S_{a_2\alpha_1}^{b_2\alpha_2}(\theta-\theta_2)\ldots
S_{a_n\alpha_{n-1}}^{b_n\alpha_n}(\theta-\theta_n)\,,
\EN
and a non-integer $\gamma_{\Phi,a}$ accounts for a semi-locality between the 
field $\Phi$ and the particle $A_a$ (see e.g. \cite{YZ}). If $z=x_1+ix_2$
and $\bar{z}=x_1-ix_2$ are complex coordinates on the plane, we say that two
fields $\Phi_1$ and $\Phi_2$ are mutually semi-local with semi-locality
index $\gamma_{\Phi_1,\Phi_2}$ if 
\EQ
\langle\cdots\Phi_1(ze^{2i\pi},\bar{z}^{-2i\pi})\Phi_2(0)\cdots\rangle=
e^{2i\pi\gamma_{\Phi_1,\Phi_2}}
\langle\cdots\Phi_1(z,\bar{z})\Phi_2(0)\cdots\rangle\,.
\label{semilocal}
\EN
Clearly, $\gamma_{\Phi_1,\Phi_2}$ is defined up to integers, and in the 
following we take it in the interval $[0,1)$.
Denoting $\varphi_a(x)$ a field which interpolates the particle $A_a$,
i.e. a field with
\EQ
\langle A_a(\theta)|\varphi_a(x)|0\rangle\neq 0\,,
\label{semilocality}
\EN
we have
\EQ
\gamma_{\Phi,a}=\gamma_{\Phi,\varphi_a}\,.
\EN
When writing (\ref{fn2}) and (\ref{fn4}) and throughout this paper we choose,
without loss of generality, the fields which interpolate the particles to be spinless. 

Since the form factors (\ref{ff}) determine by crossing all the matrix elements
of $\Phi$, and since the knowledge of all the matrix elements completely 
determines the field, the space of solutions of the equations (\ref{fn0})--(\ref{fn4})
determines the space of fields of the theory\footnote{In some cases the theory 
allows also for fields which are not simply semi-local with respect to the particles
and require a modification of (\ref{fn2}), (\ref{fn4}). In these cases ${\cal F}$, as 
defined above, is not the full space of fields. It is, however, a subspace closed 
under operator product expansion.}, which we denote by ${\cal F}$. 
Clearly, ${\cal F}$ can be decomposed into subspaces containing fields with the
same charge, spin and semi-locality indices, which are the field data 
entering the form factor equations. These subspaces are always 
infinite-dimensional.

This decomposition is conveniently characterized in the following way. 
The semi-locality property (\ref{semilocal}) can be accounted
for introducing a second additive quantity $\tilde{C}$, that we call dual 
charge, in such a way that we can write
\EQ
{\cal F}=\bigoplus_{C,\tilde{C}}{\cal F}_{C,\tilde{C}}\,,
\label{fields}
\EN
where ${\cal F}_{C,\tilde{C}}$ is the subspace containing the fields with 
charge $C$ and dual charge $\tilde{C}$. The semi-locality index entering 
(\ref{semilocal}) is given by
\EQ
\gamma_{\Phi_1,\Phi_2}=\kappa\,(C_{\Phi_1}\tilde{C}_{\Phi_2}+\tilde{C}_{\Phi_1}
C_{\Phi_2})\,,
\label{index}
\EN
with $\kappa$ a field-independent normalization for the dual charge that
we introduce for convenience. 
Additivity of the charges ensures the property $\gamma_{\Phi_1\Phi_2,\Phi_3}=
\gamma_{\Phi_1,\Phi_3}+\gamma_{\Phi_2,\Phi_3}$. We stipulate that the quanta
of the charge $C$ are measured in integer units, so that $\kappa\tilde{C}$ is
defined modulo integers like the semi-locality indices. 

The energy-momentum tensor $T^{\mu\nu}$ is neutral and local 
with respect to all the fields in the theory, and belongs to ${\cal F}_{0,0}$.
We will call ``order'' fields the fields 
belonging to ${\cal F}_{C,0}$ with $C\neq 0$, and ``disorder'' fields the 
fields belonging to ${\cal F}_{0,\tilde{C}}$ with $\tilde{C}\neq 0$. Order and 
disorder fields are mutually non-local.

If $\Phi_3$ is produced in the operator product expansion of $\Phi_1$ and 
$\Phi_2$, (\ref{semilocal}) leads to the relation $\gamma_{\Phi_1,\Phi_2}=
s_{\Phi_3}-s_{\Phi_1}-s_{\Phi_2}$. Denote by $\bar{\Phi}$ the 
charge and dual
charge conjugate of $\Phi$ ($C_\Phi+C_{\bar{\Phi}}=\tilde{C}_\Phi+\tilde{C}_
{\bar{\Phi}}=0$); $\Phi$ and $\bar{\Phi}$ have the same spin and their 
operator product expansion produces
only neutral fields, the identity among them. Then we have $\gamma_{\Phi,
\bar{\Phi}}=-2s_{\Phi}$, so that comparison with (\ref{index}) gives
\EQ
s_\Phi=\kappa\,C_\Phi\tilde{C}_\Phi+n_\Phi\,,\hspace{1cm}n_\Phi\in{\bf Z}
\label{spin}
\EN
with $n_\Phi$ accounting for the fact that the semi-locality indices are 
defined up to integers. If $s_\Phi$ is non-integer (this requires $C_\Phi$
and $\tilde{C}_\Phi$ both non-zero), $\Phi$ is said to be a 
parafermionic field \cite{FZ}, as a generalization of the fermionic fields
corresponding to half-integer spin. The field subspaces ${\cal F}_{C,
\tilde{C}}$ can further be decomposed according to the spin in the form
\EQ
{\cal F}_{C,\tilde{C}}=\bigoplus_{n\in{\bf Z}}{\cal F}^n_{C,\tilde{C}}\,,
\label{fcc}
\EN
the superscript $n$ being the integer in (\ref{spin}). For ${\cal F}_{0,0}$
$n$ is the spin itself. The subspaces ${\cal F}^n_{C,\tilde{C}}$ contain
fields corresponding to solutions of the form factor equations 
(\ref{fn0})--(\ref{fn4}) with the same values of $s_\Phi$, $C_\Phi$ and 
$\gamma_{\Phi,a}$, and are the above mentioned infinite-dimensional 
subspaces.

\resection{The operators $\Lambda_a$}
Condition (\ref{diagonal}), together with (\ref{fn1}) and (\ref{fn2}), yields
\bea
&&\lim_{\theta_n\to+\infty}F_{a_1\ldots a_n}^{\Phi}(\theta_1,\ldots,
\theta_{n-1},\theta_n+2i\pi)=
\prod_{j=1}^{n-1}e^{-2i\pi\alpha_{a_ja_n}}\lim_{\theta_n\to+\infty}
F_{a_n a_1\ldots a_{n-1}}^{\Phi}(\theta_n+2i\pi,\theta_1,\ldots,\theta_{n-1})
\nonumber\\
&&\hspace{1cm}
=e^{-2i\pi\gamma_{\Phi,a_n}}
\prod_{j=1}^{n-1}e^{-2i\pi\alpha_{a_ja_n}}\lim_{\theta_n\to+\infty}
F_{a_1\ldots a_n}^{\Phi}(\theta_1,\ldots,\theta_n)\,,
\label{limit}
\eea
and then
\EQ
F_{a_1\ldots a_n}^{\Phi}(\theta _{1},\ldots,\theta_{n})=
f_{a_1\ldots a_{n-1}}^{\Phi,a_n}(\theta _{1},\ldots,\theta_{n-1})\,
e^{y_{\Phi,a_n}\theta_n}\,,\hspace{1cm}\theta_n\to+\infty\,,\hspace{1cm}n>1
\label{Ff0}
\EN
with 
\EQ
y_{\Phi,a_n}=-\gamma_{\Phi,a_n}-\sum_{j=1}^{n-1}\alpha_{a_ja_n}+n_{\Phi,a_n}\,,
\hspace{1cm}n_{\Phi,a_n}\in{\bf Z}\,.
\label{y}
\EN
In writing (\ref{y}) we imply that, for $n>1$, the r.h.s. depends on $\Phi$ and
${a_n}$ only, a property that can be shown as follows.  
We can change the particle state while preserving (\ref{neutrality}) in two 
ways:

\noindent
i) we add/remove a particle-antiparticle pair $A_aA_{\bar{a}}$. In this case 
the product in (\ref{limit}) acquires/loses a factor
\EQ
\lim_{\theta\to-\infty}S_{aa_n}^{aa_n}(\theta)S_{\bar{a}a_n}^{\bar{a}a_n}
(\theta)=
\lim_{\theta\to-\infty}S_{aa_n}^{aa_n}(\theta)S_{{a}a_n}^{{a}a_n}
(i\pi-\theta)=1\,,
\EN
where (\ref{diagonal}) and (\ref{crossing}) have been used;

\noindent
ii) if $\Gamma_{ab}^c\neq 0$, we trade $A_c$ for $A_aA_b$, or vice versa.
In this case the product in (\ref{limit}) acquires/loses a factor
\EQ
\lim_{\theta\to-\infty}S_{aa_n}^{aa_n}(\theta)S_{ba_n}^{ba_n}(\theta)/
S_{ca_n}^{ca_n}(\theta)=1\,,
\EN
where (\ref{diagonal}) and (\ref{bootstrap}) have been used.

\noindent
This shows that the sum in (\ref{y}) depends on $\Phi$ and $a_n$ only. Finally,
also the integer $n_{\Phi,a_n}$ must have this property in such a way that the
two sides of (\ref{fn3}) and (\ref{fn4}), which relate matrix elements with a 
different number of particles, have the same limit as $\theta_n\to+\infty$.

The case $n=1$ cannot be included in (\ref{Ff0}) also because it follows from 
(\ref{fn0})
that the asymptotic limit of $F^\Phi_a(\theta)$ is determined by the spin 
$s_\Phi$, which in general does not coincide with $y_{\Phi,a}$. We now extend
(\ref{Ff0}) to the form 
\EQ
\lim_{\theta\to+\infty}e^{-y_{\Phi,a}\theta}
F_{a_1\ldots a_na}^{\Phi}(\theta _{1},\ldots,\theta_{n},\theta)=
f_{a_1\ldots a_{n}}^{\Phi,a}(\theta _{1},\ldots,\theta_{n})\,,\hspace{1cm}
n\geq 0
\label{Ff}
\EN
which for $n=0$ associates to $F_a^\Phi$ the constant $f^{\Phi,a}$. With this
additional definition we can multiply both sides of (\ref{fn0})--(\ref{fn4}) by
$e^{-y_{\Phi,a_n}\theta_n}$ and take the limit $\theta_n\to+\infty$, for any 
number of particles. The result of this operation is that the functions
$f_{a_1\ldots a_{n-1}}^{\Phi,a_n}(\theta _{1},\ldots,\theta_{n-1})$ with 
$n\geq 1$, satisfy the equations corresponding to the form factors 
$F_{a_1\ldots a_{n-1}}^{\Phi_{(a_n)}}(\theta _{1},\ldots,\theta_{n-1})$, where 
$\Phi_{(a)}$ is a field with spin, semi-locality indices and charge given by
\bea
&& s_{\Phi_{(a)}}=s_\Phi-y_{\Phi,a}\,,\label{deltaspin}\\
&& \gamma_{\Phi_{(a)},b}=\gamma_{\Phi,b}-\alpha_{ab}\,,\label{gamma}\\
&& C_{\Phi_{(a)}}=C_\Phi+C_a \label{c}\,\,.\label{deltacharge}
\eea
In other words, the limiting procedure (\ref{Ff}) defines an operation in the
space of fields which maps $\Phi$ into $\Phi_{(a)}$. Denoting by 
$\Lambda_{a}$ the corresponding operator, we can write
\EQ
\Lambda_a\Phi=\Phi_{(a)}\,,
\EN
or, on matrix elements,
\bea
\Lambda_aF_{a_1\ldots a_na}^{\Phi}(\theta _{1},\ldots,\theta_{n},\theta) &=&
\lim_{\theta\to+\infty}e^{-y_{\Phi,a}\theta}
F_{a_1\ldots a_na}^{\Phi}(\theta _{1},\ldots,\theta_{n},\theta)\nonumber\\
&=& F_{a_1\ldots a_n}^{\Phi_{(a)}}(\theta _{1},\ldots,\theta_{n})\,,
\hspace{1cm}n\geq 0\,.
\label{lambdaff}
\eea

The action of $\Lambda_a$ in (\ref{lambdaff}) yields, by construction, a finite
non-zero result for $n>0$. For $n=0$, (\ref{fn0}) and (\ref{deltaspin}) give
\EQ
\langle 0|\Phi_{(a)}(0)|0\rangle=\Lambda_aF^\Phi_a(\theta)=F^\Phi_a(0)
\lim_{\theta\to+\infty}e^{s_{\Phi_{(a)}}\theta}\,.
\label{vev}
\EN
The requirement that all fields in the theory have finite vacuum expectation 
value implies $s_{\Phi_{(a)}}\leq 0$ if $F^\Phi_a(\theta)\neq 0$, i.e.
\EQ
y_{\Phi,a}\geq s_\Phi\hspace{1cm}\mbox{if}\hspace{1cm}F^\Phi_a(\theta)\neq 0\,.
\label{lowerbound}
\EN
The vanishing of (\ref{vev}) for $s_{\Phi_{(a)}}<0$ is in agreement with 
Lorentz invariance, which prescribes that only spinless fields can have a 
non-zero vacuum expectation value.

Notice that if 
\EQ
\Phi(x)=\sum_{k=1}^Kc_k\,\Phi_k(x)
\EN
we obtain
\EQ
\Lambda_a\Phi=\left\{ 
\begin{array}{r}
c_1\,\Lambda_a\Phi_1\hspace{.5cm}\text{if}\hspace{.5cm}
y_{\Phi_1}>y_{\Phi_2}>\ldots>y_{\Phi_K}\,\,
\\ 
\\ 
\sum_{k=1}^Kc_k\,\Lambda_a\Phi_k\hspace{.5cm}\text{if}\hspace{.5cm}
y_{\Phi_1}=y_{\Phi_2}=\ldots=y_{\Phi_K}\,.
\end{array}
\right.  
\label{linearity}
\EN

\vspace{.3cm}
Using the notations
\EQ
\Theta\equiv T_\mu^\mu\,,\hspace{1cm}\Theta_a\equiv\Theta_{(a)}=
\Lambda_a\Theta\,,
\EN
(\ref{deltaspin})--(\ref{deltacharge}) give
\EQ
s_{\Theta_a}=-y_{\Theta,a}\,,\hspace{1cm}
\gamma_{\Theta_a,b}=-\alpha_{ab}\,,\hspace{1cm}
C_{\Theta_a}=C_a\,,
\label{thetaa}
\EN
where we also used the fact that the trace of the energy-momentum tensor is 
a spinless field in the Lorentz invariant theories we consider.

The spinless field $\varphi_a$ we have chosen to interpolate the particle 
$A_a$ belongs to ${\cal F}^0_{C_a,0}$. Hence, (\ref{thetaa}) and (\ref{index})
give 
\EQ
\alpha_{ab}=-\kappa\,C_a\tilde{C}_{\Theta_b}=
-\kappa\,\tilde{C}_{\Theta_a}C_b\,,
\label{alpha}
\EN
the last equality following from the symmetry of $\alpha_{ab}$ in the two
indices.
As a consequence, (\ref{gamma}) takes the form $\gamma_{\Phi_{(a)},b}=
\kappa\,(\tilde{C}_\Phi+\tilde{C}_{\Theta_a})C_b$, so that we have
\EQ
\tilde{C}_{\Phi_{(a)}}=\tilde{C}_\Phi+\tilde{C}_{\Theta_a}\,(mod\,{\bf Z}/
{\kappa})\,.
\label{ctilde}
\EN
Recalling also (\ref{y}) and (\ref{deltaspin}) we obtain
\bea
&& y_{\Phi,a}=s_\Phi-s_{\Phi_{(a)}}=-\kappa[C_a\tilde{C}_\Phi+(C_\Phi+C_a)
\tilde{C}_{\Theta_a}]+n_{\Phi,a}\,,
\label{y-charges}\\
&& n_{\Phi,a}=n_\Phi-n_{\Phi_{(a)}}\,,
\label{deltan}
\eea
so that (\ref{deltaspin})--(\ref{deltacharge}) can be restated as
\EQ
\Phi_{(a)}\in{\cal F}_{C_\Phi+C_a,\,\tilde{C}_\Phi+\tilde{C}_{\Theta_a}\,
(mod\,{\bf Z}/{\kappa})}^{n_\Phi-n_{\Phi,a}}\,\,.
\label{endo}
\EN

The fields $\Theta_a$ have spin
\EQ
s_{\Theta_a}=-\alpha_{aa}+n_{\Theta_a}\,,
\label{spintheta}
\EN
and belong to the parafermionic sector whenever $\alpha_{aa}$ is non-integer.
The asymptotic relation between the matrix elements of the 
energy-momentum tensor and those of parafermionic fields was first noted in 
\cite{Smirnov} on the basis of quantum group arguments.

For each particle $A_a$ in the theory a further decomposition of the space of 
fields according to the asymptotic behavior (\ref{Ff}) of the form
factors is obtained in the form
\EQ
{\cal F}^n_{C,\tilde{C}}=\bigoplus_{n_a\in{\bf Z}}{\cal F}^{n,n_a}_
{C,\tilde{C}}\,,
\label{fccn}
\EN
where $n_a$ is the integer in (\ref{y}). In many theories the subspaces
${\cal F}^{n,n_a}_{C,\tilde{C}}$ turn out to be finite-dimensional.

We now investigate under which conditions a spinless field is mapped into 
another spinless field by the action of the operator $\Lambda_a$. We call 
$\hat{\Omega}_{a,0}$ the field subspace containing the fields with such a 
property:
\EQ
\hat{\Omega}_{a,0}=\{\Phi\hs|\hs s_\Phi=s_{\Phi_{(a)}}=0\}\,.
\EN
Consider $\Phi\in\hat{\Omega}_{a,0}$. Since $s_\Phi=0$, $\Phi$ can only be a
linear combination of spinless fields $\Phi_i$ with $C_{\Phi_i}$ and/or 
$\tilde{C}_{\Phi_i}$ equal zero, and with $n_{\Phi_i}=0$. Similarly, 
$s_{\Phi_{i(a)}}=0$ 
implies $y_{\Phi_i,a}=n_{\Phi_i,a}=0$. At this point, (\ref{y-charges}) with
$\Phi=\Phi_i$ gives $(C_{\Phi_i}+C_a)\tilde{C}_{\Theta_a}=0$ if 
$\tilde{C}_{\Phi_i}=0$, and $(\tilde{C}_{\Phi_i}+\tilde{C}_{\Theta_a})C_a=0$ 
if $C_{\Phi_i}=0$. Hence, using $C_{\bar{a}}=-C_a$, $\tilde{C}_{\Theta_
{\bar{a}}}=-\tilde{C}_{\Theta_a}$, and recalling (\ref{endo}), we have the 
three cases\footnote{Notice that (\ref{alpha}) implies $\tilde{C}_{\Theta_a}=0$
if $C_a=0$, unless $C_b=0$ $\forall b$; but if all particles are neutral there 
is no internal symmetry, and no dual charge. Then the case $C_a=0$, 
$\tilde{C}_{\Theta_a}\neq 0$ is excluded.}
\EQ
\hat{\Omega}_{a,0}=\left\{
\begin{array}{l}
{\cal F}^{0,0}_{-C_a,0}\bigoplus{\cal F}^{0,0}_{0,-\tilde{C}_
{\Theta_a}}\,,\hspace{1.3cm} 
\mbox{if}\hs\hs C_a\neq 0,\hs\tilde{C}_{\Theta_a}\neq 0\\
\\
\bigoplus_C{\cal F}^{0,0}_{C,0}\,,\hspace{2.9cm}
\mbox{if}\hs\hs C_a\neq 0,\hs\tilde{C}_{\Theta_a}=0\\
\\
\bigoplus_C{\cal F}^{0,0}_{C,0}\,\bigoplus_{\tilde{C}}
{\cal F}^{0,0}_{0,\tilde{C}}\,,\hspace{1.3cm}
\mbox{if}\hs\hs C_a=\tilde{C}_{\Theta_a}=0\,.
\end{array}
\right.  
\label{omegahat}
\EN

The subspace $\hat{\Omega}_{a,0}$ contains in particular the spinless fields 
$\phi_a$ that $\Lambda_a$ maps onto themselves, possibly up to conjugation of 
$C$ and $\tilde{C}$. It follows from (\ref{omegahat}) and (\ref{endo}) 
that these are the solutions of the equation
\EQ
\Lambda_a\phi_a=\lambda_{\phi_a}\,\phi_{\bar{a}}\,,
\label{eigen}
\EN
with $\lambda_{\phi_a}$ constant. Denoting ${\Omega}_{a,0}$ the subspace of 
$\hat{\Omega}_{a,0}$ spanned by the solutions of (\ref{eigen}), we have
\EQ
\Omega_{a,0}\subseteq\left\{
\begin{array}{l}
{\cal F}^{0,0}_{-C_a,0}\bigoplus{\cal F}^{0,0}_{0,-\tilde{C}_
{\Theta_a}}\,,\hspace{1.1cm} 
\mbox{if}\hs\hs C_a\neq 0\\
\\ 
\bigoplus_C{\cal F}^{0,0}_{C,0}\,\bigoplus_{\tilde{C}}
{\cal F}^{0,0}_{0,\tilde{C}}\,,\hspace{1cm}
\mbox{if}\hs\hs C_a=0\,.
\end{array}
\right.  
\label{omega}
\EN

Notice that, if $C_a\neq 0$, $\phi_a$ necessarily has both a charged 
($C\neq 0$) and a neutral ($C=0$) part,
\EQ
\phi_a=\phi_a^{(c)}+\phi_a^{(n)}\,,\hspace{1cm}\text{if}\hs\hs C_a\neq 0\,.
\label{splitting}
\EN
In particular, we have $F^{\phi_a}_a\neq 0$, a condition which
holds also for $C_a=0$, unless $\phi_a$ is an order field. Equation
(\ref{vev}) yields
\EQ
\lambda_{\phi_a}=\frac{F^{\phi_a}_a}{\langle\phi_{\bar{a}}\rangle}\,,
\hspace{1cm}\text{if}\hspace{.4cm}F^{\phi_a}_a\neq 0\,
\label{eigenvalue}
\EN
for the constant in (\ref{eigen}).
Hence, for example, (\ref{eigen}) written for a particle-antiparticle state
gives
\EQ
\lim_{\theta_2\to+\infty}F^{\phi_a}_{\bar{a}a}(\theta_1,\theta_2)=
\frac{F^{\phi_a}_aF^{\phi_{\bar{a}}}_{\bar{a}}}
{\langle\phi_{\bar{a}}\rangle}\,,\hspace{1cm}\text{if}\hs\hs 
F^{\phi_a}_a\neq 0
\label{factorization}
\EN
which becomes
\EQ
\lim_{\theta_2\to+\infty}F^{\phi_a^{(n)}}_{\bar{a}a}(\theta_1,\theta_2)=
\frac{F^{\phi^{(c)}_a}_aF^{\phi^{(c)}_{\bar{a}}}_{\bar{a}}}
{\langle\phi^{(n)}_{\bar{a}}\rangle}\,,\hspace{1cm}\text{if}\hs\hs C_a\neq 0\,.
\label{cn}
\EN

We will see in the next section that equation (\ref{eigen}) normally admits 
solutions, so that $\Omega_{a,0}$ is non-empty and contains fields whose 
form factors have the mildest asymptotic behavior which (\ref{lowerbound}) 
allows for spinless fields ($y_{\Phi,a}=0$). In this sense the fields $\phi_a$
can be called ``fundamental'', meaning that for any positive $k$ there exist 
spinless fields $\Phi$ with the same charge and dual charge content as 
$\phi_a$, whose
matrix elements are, for this reason, solutions of exactly the same form factor
equations, with a different asymptotic behavior, $y_{\Phi,a}=n_{\Phi,a}=k$. 
We denote $\Omega_{a,k}$ the space spanned by such fields and use the notation
\EQ
\Omega_{a}=\bigoplus_{k=0}^\infty\Omega_{a,k}
\label{o}
\EN
for the whole sector of the space of fields related, in the way we just 
explained, to the fundamental fields $\phi_a$.
The derivatives $(\partial\bar{\partial})^k\phi_a$, whose form factors are 
immediately obtained from those of $\phi_a$, are examples of fields 
belonging to $\Omega_{a,k}$. We know that the action of $\Lambda_a$ on a 
field in $\Omega_{a,k}$ gives a field with spin $-k$. 

The structure of $\Omega_{a}$, with its internal gradation in terms of a
non-negative integer $k$, is clearly reminiscent of the characterization of 
the space of fields in conformal field theory \cite{BPZ}. This is not totally
unexpected once we consider that, up to symmetry breaking effects, the field
content of the massive theory and that of its massless (conformal) limit should
coincide. Let us recall that in a two-dimensional conformal field theory the
space of fields is the direct sum of ``families'' corresponding to lowest 
weight representations of the infinite-dimensional Virasoro algebra. All the 
fields within a family behave in the same way under the internal symmetries
of the theory, namely, for the cases of interest in this paper, have the same
charge and dual charge. Each family consists of a ``primary'' field
and infinitely many ``descendants'' whose scaling dimension differs from that
of the primary by a positive integer. Restricting our attention to spinless
primaries and spinless descendants, the scaling dimension of a primary can
be denoted $2\Delta$ and that of a descendant $2(\Delta+l)$, where $l$ is 
called level of the descendant; there exist descendants for all positive 
values of $l$. If we denote by $V_{\Delta,l}$ the space of spinless descendants
of level $l$, the restriction to spinless fields of a family associated to a 
spinless primary can be written as
\EQ
V_\Delta=\bigoplus_{l=0}^\infty V_{\Delta,l}\,.
\label{v}
\EN

Considering that the derivative $(\partial\bar{\partial})^l$ of the primary
belongs to $V_{\Delta,l}$, the hypothesis that $k$ in (\ref{o}) and $l$ in
(\ref{v}) play the same role is very natural. We refrain however from a 
straightforward identification between $k$ and $l$. If it is reasonable to 
expect that
\EQ
\Omega_{a}=\bigoplus_{\Delta\in{\cal D}_a}V_{\Delta}\,,
\label{omegav}
\EN
with the sum taken over the families having the same charge and dual charge
content as $\phi_a$ \footnote{Consistence of this statement requires
\EQ
\Omega_{a}=\Omega_{b}\hspace{.5cm}\text{if}\hspace{.5cm}C_a=C_b\,.
\label{samecharge}
\EN}, 
the hypothesis that $\bigoplus_{\Delta\in{\cal D}_a}V_{\Delta,l}$ coincides
with $\Omega_{a,l}$ is probably too strong in the general case. Indeed, it is
difficult to exclude that, for some $\Delta\in{\cal D}_a$, $V_{\Delta,l}
\subseteq\Omega_{a,k}$ with a fixed positive value of $k-l$.

Even with this caveat, the space $\Omega_{a,0}$ should be spanned by 
primary fields\footnote{In the massive context we call primary field a field 
which becomes a conformal primary in the massless limit.}. On the other hand, 
by definition, a basis in $\Omega_{a,0}$ is provided by the solutions of 
equation (\ref{eigen}). Whether, when $\text{dim}\,\Omega_{a,0}>1$, the basis 
of solutions of (\ref{eigen}) coincides with the basis of primary fields,
is an essential question. Also, when $\phi_a$ splits into two
components as in (\ref{splitting}), the simplest 
expectation is that these components correspond to primary fields with the 
same scaling dimension related by a symmetry transformation. 

In the next section we illustrate how this scenario appears to be supported
by the results so far available for several models. Before that, let 
us recall that, for reflection positive theories\footnote{Reflection positivity
requires that the two-point correlation function of any field other than the 
identity is positive and monotonically decreasing with distance. In particular,
it excludes negative scaling dimensions.}, it was shown in \cite{immf} that
\EQ
y_{\Phi,a}\leq\Delta_\Phi\hspace{1cm}\text{for}\hspace{1cm}s_\Phi=0\,,
\label{upperbound}
\EN
where $2\Delta_\Phi$ is the scaling dimension of $\Phi$. Relevant (in the 
renormalization group sense) fields have $\Delta_\Phi<1$, and in a reflection
positive theory are necessarily primaries. In practice (\ref{upperbound}) 
turns out to be extremely constraining for the purpose of the identification
of these fields. The lower bound (\ref{lowerbound}) helps understanding this
circumstance on general grounds.

\resection{Application to models}
\subsection{${\bf Z}_N$ models}
For an internal symmetry group $G={\bf Z}_N$, $N\geq 2$, 
the simplest (minimal) solution of the
equations (\ref{unitarity})--(\ref{yb}) is a scattering theory \cite{KS} of 
particles $A_a$, $a=1,\ldots,N-1$, with masses 
\EQ
m_a=2m\,\sin\frac{\pi a}{N}\,,
\label{znmasses}
\EN
${\bf Z}_N$ charges $C_a=a$, and scattering amplitudes 
\bea
&& S_{ab}^{cd}(\theta)=\delta_a^c\delta_b^dS_{ab}(\theta)\,,
\label{reflectionless}\\
&& S_{ab}(\theta)=\prod_{l=0}^{a-1}\prod_{m=0}^{b-1}f_{2/N}\left(\theta+
\frac{i\pi}{N}(a-2l)-\frac{i\pi}{N}(b-2m)\right)\,,\\
&& f_\alpha(\theta)=\frac{\sinh\frac12(\theta+i\pi\alpha)}{
\sinh\frac12(\theta-i\pi\alpha)}\,\,.
\eea

Since $f_\alpha(\theta)\to e^{\pm i\pi\alpha}$ as $\theta\to\pm\infty$, the
asymptotic phases obtained from (\ref{diagonal}) are
\EQ
\alpha_{ab}=\frac{ab}{N}\,.
\EN
Chosing $\kappa=-1/N$ and comparing with (\ref{alpha}) we obtain 
\EQ
\tilde{C}_{\Theta_a}=C_a=a\,.
\EN
The fact that $\kappa\tilde{C}$ is defined modulo integers means that 
$\tilde{C}$ is identified modulo $N$, so it is the charge associated to a 
group that we can denote $\tilde{Z}_N$. As a consequence, the sum in 
(\ref{fields}) runs over integer values from $0$ to $N-1$ for both $C$ and 
$\tilde{C}$. The fields $\Theta_a$ with spin 
\EQ
s_{\Theta_a}=-\frac{a^2}{N}+n_{\Theta_a}\,,
\label{spinzn}
\EN
are parafermions belonging to ${\cal F}_{a,a}$, and appear in the 
operator product expansion of the order fields in ${\cal F}_{a,0}$ with the 
disorder fields in ${\cal F}_{0,a}$.

These conclusions about the space of fields associated to the minimal 
${Z}_N$--invariant scattering theories fully agree with the known fact that 
these theories describe the scaling limit of $Z_N$-invariant lattice 
models \cite{FZ,ABF}, and are obtained as the symmetry 
preserving perturbations of the ${\bf Z}_N$-invariant conformal field theories 
with central charge $2(N-1)/(N+2)$ constructed in \cite{FZ}. These 
theories indeed exhibit a space of fields classified in terms of $Z_N\times
\tilde{Z}_N$ charges. In particular, the spin of the fields in the 
parafermionic sector is known to be $\pm a^2/N$ plus integers, and includes 
(\ref{spinzn}). The ``order parameters'' $\sigma_a$, $a=1,\ldots,N-1$ (which in
our notation are the most relevant fields in ${\cal F}_{a,0}^0$) have scaling
dimensions $a(N-a)/N(N+2)$, as well as the ``disorder parameters''
$\mu_a$ (the most relevant fields in ${\cal F}_{0,a}^0$). Since the fields 
$\sigma_a$ and $\mu_a$ are exchanged by duality trasformations, their relative
normalization is fixed by the condition
\EQ
\lim_{|x|\to 0}\frac{
\langle\sigma_a(x)\sigma_{N-a}(0)\rangle}{\langle\mu_a(x)\mu_{N-a}(0)\rangle}=1
\,.
\label{sigmamu}
\EN

Following the reasonings of the previous section, the expectation is that in 
this case the solutions of (\ref{eigen}) decompose as in (\ref{splitting}) with
\EQ
\phi_a^{(c)}=\sigma_{N-a}\,,\hspace{1cm}\phi_a^{(n)}=b_a\,\mu_{N-a}\,,
\label{phizn}
\EN
with $b_a$ a constant. 

Form factor solutions have been studied in \cite{BKW,YZ} for $N=2$ (Ising
model) and in \cite{Smirnovpotts,DC} for $N=3$ (three-state Potts model). It 
can be checked from these results that a solution to (\ref{eigen}) exists and
is unique (up to normalization) for a given $a$; moreover (\ref{phizn}) holds 
and $|\lambda_{\phi_a}|=|F^{\sigma_{N-a}}_a/b_{N-a}\langle\mu_a\rangle|$ equals
$1$ for $N=2$, $a=1$, and $0.9839..$ for $N=3$, $a=1,2$.

In principle, the normalization constants $b_a$ can be determined exactly 
through a resummation of the form factor spectral decompositions of the 
correlators entering (\ref{sigmamu}). In practice, this can be 
achieved only for $N=2$ and gives $|b_1|=1$ (see e.g. \cite{YZ}). For $N=3$ the
result $|b_1|=|b_2|=1$ was obtained within $1\%$ accuracy in \cite{CDGJM} 
through comparison with high precision Monte Carlo data for the lattice model.

It would be interesting to generalize the analysis to $N>3$ exploiting the
form factor formulae obtained in \cite{BFK,FPP}. 

\subsection{Sine-Gordon model}
The sine-Gordon model is the integrable quantum field theory defined by the 
action
\EQ
{\cal A}_{SG}=\int d^2x\,\left(\frac{1}{2}\,(\partial_\mu\varphi)^2-
\tau\cos\beta\varphi\right)\,.
\label{sg}
\EN
The elementary excitations are the solitons $A_+$ and anti-solitons $A_-$ 
which interpolate between adjacent degenerate minima of the periodic 
potential. Hence, the particles $A_a$, $a=\pm 1$, carry a topologic charge
$a$. 

It is well known \cite{Coleman} that the sine-Gordon model is equivalent 
through fermionization to the massive Thirring model, the theory of a Dirac
fermion with four-spin interaction which explicitly exhibits invariance
(hidden in (\ref{sg})) under a group $G=U(1)$. The solitons of the bosonic
action are the Thirring fermions, and the topologic charge is the $U(1)$ 
charge: $C_a=a$.

The scattering of solitons and antisolitons is determined by the amplitudes 
\cite{ZZ}
\bea
&& S_{++}^{++}(\theta)=S_{--}^{--}(\theta)=
S_0(\theta)=-\exp\left\{-i\int_0^\infty\frac{dx}{x}
\frac{\sinh\frac{x}{2}\left(1-
\frac{\xi}{\pi}\right)}{\sinh\frac{x\xi}{2\pi}\cosh\frac{x}{2}}
\sin\frac{\theta x}{\pi}\right\}\,,\\
&& S_{+-}^{+-}(\theta)=S_{-+}^{-+}(\theta)=
-\frac{\sinh\frac{\pi\theta}{\xi}}{\sinh\frac{\pi}{\xi}(\theta-i\pi)}
S_0(\theta)\,,\label{saforward}\\
&& S_{+-}^{-+}(\theta)=S_{-+}^{+-}(\theta)=
-\frac{\sinh\frac{i\pi^2}{\xi}}{\sinh\frac{\pi}{\xi}(\theta-i\pi)}
S_0(\theta)\,,\label{saback}\\
&& \hspace{3cm}\xi=\frac{\pi\beta^2}{8\pi-\beta^2}\,\,.
\eea
The asymptotic diagonality condition (\ref{diagonal}) is satisfied with 
\EQ
\alpha_{ab}=-\frac14\left(1+\frac{\pi}{\xi}\right)ab=-\frac{2\pi}{\beta^2}ab\,,
\hspace{1cm}a,b=\pm 1\,.
\label{alphasg}
\EN
Comparison with (\ref{alpha}) (with the choice $\kappa=1$) and 
(\ref{spintheta}) gives
\bea
&& \tilde{C}_{\Theta_a}=\frac{2\pi}{\beta^2}a\,(mod\,{\bf Z})\,,\\
&& s_{\Theta_a}=\frac{2\pi}{\beta^2}+n_{\Theta_a}\,.
\eea
Since $\beta$ is a continuous parameter the dual charge is not quantized, and
the sum in (\ref{fields}) runs over all integers for $C$, and over the 
interval $[0,1)$ for $\tilde{C}$. 

The charges $C$ and $\tilde{C}$ can indeed be identified for the primary fields
of the sine-Gordon model. These can be written as (see e.g. \cite{ZZreport})
\EQ
U_{m,\nu}(x)=e^{i\left[\frac{2\pi}{\beta}m\tilde{\varphi}(x)+\nu\beta\varphi(x)\right]}\,,
\label{u}
\EN
where $m\in{\bf Z}$ is the $U(1)$ charge. 
At the gaussian fixed point the bosonic field can be written as $\varphi(x)=
\phi(z)+\bar{\phi}(\bar{z})$, and $\tilde{\varphi}(x)\equiv\phi(z)-
\bar{\phi}(\bar{z})$. The fields (\ref{u}) have semi-locality indices
and spin given by
\bea
&& \gamma_{(m_1,\nu_1),(m_2,\nu_2)}=m_1\nu_2+m_2\nu_1\,,\\
&& s_{m,\nu}=m\nu\,.
\eea
Recalling (\ref{index}) and (\ref{spin}), we have
\EQ
C_{U_{m,\nu}}=m\,,\hspace{1cm}\tilde{C}_{U_{m,\nu}}=\nu\,(mod\,{\bf Z})\,.
\EN
The fields $U_{m,0}$ and
$U_{0,\nu}$ are spinless and have scaling dimensions \cite{ZZreport} 
$\pi m^2/\beta^2$ and $\nu^2\beta^2/4\pi$, respectively. In particular, the 
trace of the energy-momentum tensor $\Theta\sim\cos\beta\varphi\sim 
U_{0,1}+U_{0,-1}$ belongs to ${\cal F}_{0,0}^0$, as it should, and has scaling 
dimension $\beta^2/4\pi$.

From the first line of (\ref{omega}) we expect that the solutions of (\ref{eigen}) 
are
\EQ
\phi_\pm=U_{\mp 1,0}+b_\pm\,U_{0,\mp 2\pi/\beta^2}= 
e^{\mp i\frac{2\pi}{\beta}\tilde{\varphi}}+b_\pm\,e^{\mp i\frac{2\pi}{\beta}\varphi}\,,
\label{phipm}
\EN
with $b_\pm$ constants. Observe that the fields in the linear combination,
which we select on these grounds, have indeed the same scaling dimension 
$\pi/\beta^2$. Moreover, if the results for the form factors of the fields (\ref{u}) 
given in \cite{Lukyanov,LZ} are not sufficient to check that (\ref{phipm}) 
satisfy (\ref{eigen}), one can check that $U_{0,-2\pi a/\beta^2}$ indeed 
belongs to ${\cal F}_{0,-2\pi a/\beta^2}^{0,n_a}$ with $n_a=0$. 

For $\beta^2<4\pi$ the soliton-antisoliton interaction is attractive and the 
amplitudes (\ref{saforward}), (\ref{saback}) possess simple 
poles at $\theta=i(\pi-n\xi)$ corresponding to bound states (breathers) 
$B_n$ with masses
\EQ
m_n=2M\sin\frac{n\xi}{2}\,,\hspace{1cm}1\leq n<\left[\frac\pi\xi\right]
\label{mn}
\EN
where $M$ is the soliton mass and $[x]$ denotes the integer
part of $x$. The lightest breather $B_1$ is the particle interpolated by the
boson $\varphi$. Since $C_{B_n}=0$, we also have $\tilde{C}_
{\Theta_{B_n}}=0$ \footnote{One can check that the soliton-breather and
breather-breather scattering amplitudes \cite{ZZ} go asymptotically to $1$.}. 
In this case the space $\Omega_{B_n,0}$ of solution of the equation 
\EQ
\Lambda_{B_n}\phi_{B_n}=\lambda_{\phi_{B_n}}\,\phi_{B_n}
\label{lambdabn}
\EN
is expected to contain all the fields (\ref{u}) with zero spin (i.e. with $m$ or
$\nu$ equal zero). The form factor results of \cite{KM,Lukyanov,LukyToda} 
indicate that the fields $U_{0,\alpha}$ satisfy (\ref{lambdabn}) and that
\EQ
\lambda_{(U_{0,\alpha})_{B_1}}=-2i\cos\frac{\xi}{2}\sqrt{2\sin\frac{\xi}{2}}\,
\exp\left[-\int_0^\xi\frac{dt}{2\pi}\frac{t}{\sin t}\right]\,
\frac{\sin\xi\alpha}{\sin\xi}\,.
\EN

\subsection{Theories without internal symmetries}
In an integrable theory without internal symmetries all particles are 
neutral, there is no mass degeneracy and the scattering is necessarily 
diagonal. Hence (\ref{diagonal}) is satisfied and $\alpha_{ab}=0$ $\forall 
a,b$, as implied by (\ref{alpha}) and $C_a=0$ $\forall a$. This agrees with
the known fact that in these cases all the amplitudes can be expressed as
\EQ
S_{ab}(\theta)=\prod_{\gamma\in{\cal G}_{ab}}t_\gamma(\theta)\,,
\label{asymm}
\EN
\EQ
t_\gamma(\theta)=\frac{\tanh\frac12(\theta+i\pi\gamma)}
                      {\tanh\frac12(\theta-i\pi\gamma)}\,\,.
\EN

Since $C_\Phi=\tilde{C}_\Phi=0$ $\forall \Phi$, it follows from (\ref{spin})
that all the fields have integer spin and that (\ref{fields}) and (\ref{fcc}) 
reduce to the sum
\EQ
{\cal F}=\bigoplus_{s\in{\bf Z}}{\cal F}_{0,0}^s
\EN
over subspaces with spin $s$. Such subspaces are further decomposed as in
(\ref{fccn}) according to the value of $n_{\Phi,a}=y_{\Phi,a}$. The solutions
of (\ref{eigen}) all belong to ${\cal F}_{0,0}^{0,0}$ and have 
$\lambda_{\phi_a}=F_a^{\phi_a}/\langle\phi_a\rangle$. Equation (\ref{omegav})
now becomes
\EQ
\Omega_{a}={\cal F}_{0,0}^{0}=\bigoplus_{\Delta}V_{\Delta}\,,
\label{nosymmetry}
\EN
where the value of $a$ is immaterial, as required by (\ref{samecharge}), and
the sum runs over all values of $\Delta$ allowed in the theory. For this kind 
of theories, equation (\ref{eigen}) amounts to a particular case of the 
asymptotic factorization property for form factors proposed in \cite{DSC}
and generalized in \cite{ttbar}.

\vspace{.3cm}
A particularly interesting example within this class of theories is the scaling
Ising model with magnetic field at critical temperature \cite{Taniguchi}. The 
$S$-matrix bootstrap has initial condition (\ref{asymm}) with ${\cal G}_{11}=
\{2/3,2/5,1/15\}$ and closes on eight species of particles $A_a$, $a=1,\ldots,
8$, with different masses. Results for the form factors in this theory have 
been
obtained in \cite{immf,DS,DGM} and are reviewed in \cite{report}. The evidence
is that equation (\ref{eigen}) admits two solutions, the same for all values
of $a$, in agreement with (\ref{samecharge}) and with the fact that the Ising 
model possesses two non-trivial (i.e. other than the identity) primary fields, 
the spin $\sigma$ and the energy $\varepsilon$. In particular, for $a=1$ one 
obtains
\EQ
\lambda_{\phi_1}=\left\{
\begin{array}{l}
-0.640902..\\ 
\\
-3.70658..\,.
\end{array}
\right.  
\label{immf}
\EN
The form factor solution corresponding to the field $\sigma$ can be easily
identified because in this model $\sigma\sim\Theta$, and yields a ratio
$F_1^{\sigma}/\langle\sigma\rangle$ which coincides with the upper value in
(\ref{immf}). On the other hand, numerical estimates on the lattice model
give $|F_1^{\sigma}/\langle\sigma\rangle|=0.6408(3)$ and 
$|F_1^{\varepsilon}/\langle\varepsilon\rangle|=3.707(7)$ \cite{CH}, providing 
substantial evidence that (\ref{eigen}) correctly selects also the second 
primary field.

Infinite series of integrable theories without internal symmetries are
obtained from massive perturbations of the non-reflection-positive conformal
minimal models ${\cal M}_{2,2m+3}$, $m=1,2,\ldots$, with central charge
$1-3(2m+1)^2/(2m+3)$. The case $m=1$ corresponds to the Lee-Yang model
describing the edge singularity of the zeros of the partition function of the
Ising model in an imaginary magnetic field \cite{YL}--\cite{CMyanglee}.
The form factor results of \cite{Alyosha} for $m=1$ and \cite{AMV} for 
$m=2,3$ indicate that the above mentioned asymptotic factorization property,
and consequently equation (\ref{eigen}), admit $m$ solutions, in perfect 
agreement with the number of non-trivial primary fields present at criticality.

For the Lee-Yang model the isomorphism between the conformal space 
of fields and that determined by the form factor equations has been shown in 
\cite{isom,seven}. It gives, in particular,
\EQ
\Omega_{a,l}={\cal F}_{0,0}^{0,l}=\bigoplus_{\Delta}V_{\Delta,l}\,,
\label{omegal}
\EN
with the sum running over the two values allowed in the theory, $\Delta=0$ 
(the identity family\footnote{Since the identity cannot be considered as a 
solution of (\ref{eigen}), this trivial primary should be added to the l.h.s.
of (\ref{omegal}) for $l=0$.}) and $\Delta=-1/5$. This shows that for the 
Lee-Yang model
the identification between $k$ in (\ref{o}) and $l$ in (\ref{v}) is complete.

\resection{Conclusion}
In this paper we identified a number of general facts concerning
the classification of quantum fields in integrable theories with
asymptotically diagonal scattering and additive charges. An essential 
role is played by
the asymptotic behavior of form factors, which allows for the 
introduction of operators $\Lambda_a$ mapping fields into fields.
A notion of massive primary field is naturally associated to these 
operators through equation (\ref{eigen}). There is evidence for 
several models that the fields selected in this way become conformal 
primaries in the massless limit. The possibility of classifying 
scalar fields into subspaces labelled by a non-negative integer 
indicates that the level structure implied by conformal symmetry at
fixed points is recovered starting from particle dynamics
away from criticality.

According to (\ref{samecharge}), operators $\Lambda_a$ associated to
particles with the same charge are expected to 
select the same sector of the space of fields. Generically, however, 
not all values of charge are realized by single-particle 
states. For example, the ${\bf Z}_N$ models do not possess neutral
particles, while in sine-Gordon there are no particles with charge
larger than $1$ in absolute value. It seems reasonable to expect 
that for these cases the corresponding sectors of field space are 
associated to operators of the form $\Lambda_{a_1}\ldots
\Lambda_{a_n}$, with $\sum_{i=1}^nC_{a_i}$ equal to the 
desired charge value. So, as an example, the field $\Theta$ in
the ${\bf Z}_N$ models would be solution of the equation
$\Lambda_a\Lambda_{\bar{a}}\phi=\lambda\phi$.

The generalization of our analysis to integrable theories outside
the class we considered here is also an interesting issue\footnote{See
\cite{BW} for results on the asymptotic properties of form
factors in the $O(n)$ non-linear sigma model with $n>2$.}. It seems plausible 
that equations of type (\ref{lambdaff}) can serve as a definition for
operators $\Lambda_a$ in more general integrable theories, or even in absence 
of integrability. Speculations about higher dimensions are also possible.

\vspace{1cm}\textbf{Acknowledgments.} Work supported in part by the ESF 
grant INSTANS.


\end{document}